\documentclass[prb,preprintnumbers,amsmath,amssymb,floatfix]{revtex4}

\usepackage{graphicx}
\usepackage{subfigure}
\usepackage{dcolumn}

\makeatletter

\begin{document}
\baselineskip 16pt

\title{Quantum Theory Approach for Neutron Single and Double-Slit Diffraction}
\author{Xiang-Yao Wu$^{a}$ \footnote{E-mail: wuxy2066@163.com},
Bai-Jun Zhang$^{a}$, Zhong Hua$^{a}$\\Xiao-Jing Liu$^{a}$, Yi-Heng
Wu$^{a}$, Hou-Li Tang$^{a}$ and JIng-Wu Li$^{b}$ }

\affiliation{a.Institute of Physics, Jilin Normal University,
Siping 136000, China\\b.Institute of Physics, Xuzhou Normal
University, Xuzhou 221000, China}

\begin{abstract}
We provide a quantum approach description of neutron single and
double-slit diffraction, with specific attention to the cold
neutron diffraction ($\lambda \approx 20$\AA) carried out by
Zeilinger et al. in 1988. We find the theoretical results
are good agreement with experimental data.\\
\vskip 5pt


\end{abstract}
\maketitle

\maketitle {\bf 1. Introduction} \vskip 8pt

The matter-wave diffraction has become a large field of interest
over the last years, and it is extended to electron, neutron,
atom, more massive, complex objects, like large molecules $I_{2}$,
$C_{60}$ and $C_{70}$, which were found in experiments [1-5]. At
present, There are classical and quantum methods to study
interference and diffraction [6-12]. As is well known, the
classical optics with its standard wave-theoretical methods and
approximations, in particular those of Huygens and Kirchhoff, has
been successfully applied to classical optics, and has yielded
good agreement with many experiments. This simple wave-optical
approach also gives a description of matter wave diffraction.
However, matter-wave interference and diffraction are quantum
phenomena, and its full description needs quantum mechanical
approach. Recently, there are some quantum theory approach to
study electron and neutron diffraction, and obtain some important
and new results [13-17]. In viewpoint of quantum mechanics, the
neutron has the wave nature, which is described by wave function
$\psi(\vec{r},t)$, and the wave function $\psi(\vec{r},t)$ has
statistical meaning, i.e., $\mid\psi(\vec{r},t)\mid^{2}$ can be
explained as particle's probability density. For the single and
double-slit diffraction, if we can calculate the neutron wave
function $\psi(\vec{r},t)$ distributing on display screen, then we
can obtain the diffraction intensity, since the diffraction
intensity is directly proportional to
$\mid\psi(\vec{r},t)\mid^{2}$. In the single and double-slit
diffraction, the neutron wave functions can be divided into three
parts. The first is the incident area, and the neutron wave
function is a plane wave. The second is the slit area, where the
neutron wave function can be calculated by the Schr\"{o}dinger
wave equation. The third is the diffraction area, where the
neutron wave function can be obtained by Kirchhoff's law.
Otherwise, we consider the decoherence effect in the double slit
diffraction. We know decoherence is the irreversible emergence of
classical properties when an isolated system interacts with an
environment [18]. The environment can be constituted by many
randomly distributed particles interacting with the system by
means of scattering processes. When these events occur in a large
number, the off-diagonal elements of the system reduced density
matrix undergo an exponential damping [19], this making the system
to quickly lose its coherence, i.e., the decoherence is the
dynamic suppression of the interference terms owing to the
interaction between system and environment. In this paper, we
study the neutron single and double-slit diffraction with the
quantum approach, and analyze the influence of the decoherence
machanism to the double slit diffraction. We compare our
calculation results to the cold neutron  ($\lambda \approx 20$\AA)
diffraction experiment carried out by Zeilinger et al. in 1988
[20]. We find the decoherence machanism has improved the
calculation result of the double slit diffraction, and the theory
results are agreement with the experiment data.
 \vskip 5pt
\newpage
 \setlength{\unitlength}{0.1in}
\begin{center}
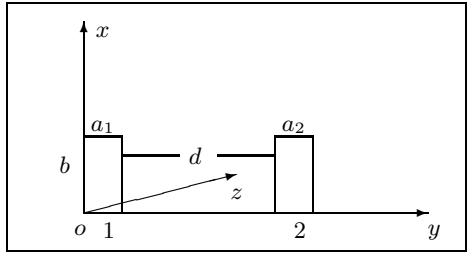
\begin{figure}
\begin{picture}(100,15)

\put(26,4){\vector(1,0){18}}
\put(26,4){\vector(0,1){10}}
\put(26,4){\vector(4,1){8}}

 \put(22,2){\line(1,0){24}}
 \put(22,2){\line(0,1){13}}
 \put(46,2){\line(0,1){13}}
 \put(22,15){\line(1,0){24}}

 \put(26,8){\line(1,0){2}}
 \put(28,4){\line(0,1){4}}
\put(36,4){\line(0,1){4}}
 \put(36,8){\line(1,0){2}}
 \put(38,4){\line(0,1){4}}

 \put(28,7){\line(1,0){3}}
 \put(33,7){\line(1,0){3}}

 \put(44,2.5){\makebox(2,1)[l]{$y$}}
 \put(26,13){\makebox(2,1)[c]{$x$}}
 \put(25.5,2.6){\makebox(2,1)[l]{$o$}}
 \put(33,4.5){\makebox(2,1)[c]{$z$}}
 \put(24,6){\makebox(2,1)[c]{$b$}}
 \put(26,8){\makebox(2,1)[c]{$a_{1}$}}
 \put(36,8){\makebox(2,1)[c]{$a_{2}$}}
 \put(31.5,6.5){\makebox(2,1)[l]{$d$}}
 \put(27,2.6){\makebox(2,1)[l]{$1$}}
 \put(37,2.6){\makebox(2,1)[l]{$2$}}
\end{picture}
\caption{Double-slit geometry with $a_{1}$ the first slit width,
$a_{2}$ the second slit width, $b$ the slit length and $d$ the
distance between the two slits.} \label{moment}
\end{figure}
\end{center}
{\bf 2. Quantum approach of neutron diffraction}

\vskip 8pt
 In an infinite plane, we consider a double-slit, its
width $a_1$ and $a_2$, length $b$, thickness $c$ and the
slit-to-slit distance $d$ are shown in FIG.\,1. The $x$ axis is
along the slit length and the $y$ axis is along the slit width. We
calculate the neutron wave function in the first single slit
(left) with the Schr\"odinger equation, and the neutron wave
function of the second single-slit (right) can be obtained easily.
At time $t$, we suppose that the incident plane wave travels along
the $z$ axis. It is
\begin{equation}
\psi_{0}(z, t)=Ae^{\frac{i}{\hbar}(pz-Et)},
\end{equation}
where $A$ is plane wave amplitude.

The potential in the first single slit is
\begin{eqnarray}
V(x,y,z)= \left \{ \begin{array}{ll}
   0  \hspace{0.3in} 0\leq x\leq b, 0\leq y\leq a_{1}, 0\leq z\leq c, \\
   \infty  \hspace{0.3in}  otherwise,
   \end{array}
   \right.
\end{eqnarray}
where $c$ is the thickness of the single slit. The time-dependent
and time-independent Schr\"odinger equations are
\begin{equation}
i\hbar\frac{\partial}{\partial
t}\psi(\vec{r},t)=-\frac{\hbar^{2}}{2M}(\frac{\partial^{2}}{\partial
x^{2}}+\frac{\partial^{2}}{\partial
y^{2}}+\frac{\partial^{2}}{\partial z^{2}})\psi(\vec{r},t),
\end{equation}
\begin{equation}
\frac{\partial^{2}\psi(\vec{r})}{\partial
x^{2}}+\frac{\partial^{2}\psi(\vec{r})}{\partial
y^{2}}+\frac{\partial^{2}\psi(\vec{r})}{\partial
z^{2}}+\frac{2ME}{\hbar^{2}}\psi(\vec{r})=0,
\end{equation}
where $M(E)$ is the mass(energy) of the neutron. The relation
between $\psi(\vec{r},t)$ and $\psi(\vec{r})$ is
\begin{equation}
\psi(x,y,z,t)=\psi(x,y,z)e^{-\frac{i}{\hbar}Et}.
\end{equation}

In Eq. (4), the wave function $\psi(x,y,z)$ satisfies the boundary
conditions
\begin{equation}
\psi(0,y,z)=\psi(b,y,z)=0,
\end{equation}
\begin{equation}
\psi(x,0,z)=\psi(x,a_{1},z)=0.
\end{equation}
The Eq. (4) can be solved by the method of separation of variable.
By writing

\begin{equation}
\psi(x,y,z)=X(x)Y(y)Z(z).
\end{equation}
The general solution of Eq. (3) is
\begin{eqnarray}
\psi_{1}(x,y,z,t)&=&\sum_{mn}\psi_{mn}(x,y,z,t) \nonumber\\
&=&\sum_{mn}D_{mn}\sin{\frac{n\pi x}{b}}\sin{\frac{m\pi
y}{a_{1}}}e^{i\sqrt{\frac{2ME}{\hbar^{2}}-\frac{n^{2}\pi^{2}}{b^{2}}-\frac{m^{2}\pi^{2}}{a_{1}^{2}}}z}e^{-\frac{i}{\hbar}Et}.
\end{eqnarray}
Eq. (9) is the neutron wave function in the first single slit.
Since the wave functions are continuous at $z=0$, we have
\begin{equation}
\psi_{0}(x,y,z,t)\mid_{z=0}=\psi_{1}(x,y,z,t)\mid_{z=0}.
\end{equation}
From Eqs. (2), (6) and (9), we can obtain the Fourier coefficient
$D_{mn}$ by Fourier transform
\begin{eqnarray}
D_{mn}&=&\frac{4}{a_{1}
b}\int^{a_{1}}_{0}\int^{b}_{0}A\sin{\frac{n\pi
\xi}{b}}\sin{\frac{m\pi \eta}{a_{1}}}d\xi d\eta \nonumber\\
&=&\left \{ \begin{array}{ll}
   \frac{16A}{mn\pi^{2}} \hspace{0.6in} m,n, odd, \\
   0 \hspace{0.9in} otherwise,
   \end{array}
   \right.
\end{eqnarray}
substituting Eq. (11) into Eq. (9), we can obtain the neutron wave
function in the first single slit.

\begin{eqnarray}
\psi_{1}(x,y,z,t)&=&\sum_{m,n=0}^{\infty}\frac{16A}{(2m+1)(2n+1)\pi^{2}}\sin{\frac{(2n+1)\pi
x}{b}}\sin{\frac{(2m+1)\pi y}{a_{1}}} \nonumber\\&& \cdot
 e^{i\sqrt{\frac{2ME}{\hbar^{2}}-\frac{(2n+1)^{2}\pi^{2}}{b^{2}}
-\frac{(2m+1)^{2}\pi^{2}}{a_{1}^{2}}}z}e^{-\frac{i}{\hbar}Et}.
\end{eqnarray}

The neutron wave function in the second single slit can be
obtained by making the coordinate translations $x'=x$,
$y'=y-a_{1}-d$, $z'=z$, and we can obtain the neutron wave
function $\psi_{2}(x,y,z,t)$ in the second slit
\begin{eqnarray}
\psi_{2}(x,y,z,t)&=&\sum_{m,n=0}^{\infty}\frac{16A}{(2m+1)(2n+1)\pi^{2}}
\sin{\frac{(2n+1)\pi x}{b}}\sin{\frac{(2m+1)\pi
(y-a_{1}-d)}{a_{2}}} \nonumber\\&& \cdot
 e^{i\sqrt{\frac{2ME}{\hbar^{2}}-\frac{(2n+1)^{2}\pi^{2}}{b^{2}}
-\frac{(2m+1)^{2}\pi^{2}}{a_{2}^{2}}}z}e^{-\frac{i}{\hbar}Et}.
\end{eqnarray}
\vskip 8pt

{\bf 3. The wave function of neutron diffraction} \vskip 8pt With
Kirchhoff's law, we can calculate the neutron wave function in the
diffraction area. It can be calculated by the formula [21]
\begin{equation}
\psi_{out}({\pmb r},t)=-\frac{1}{4\pi}\int_{s}\frac{e^{ikr}}{r}
{\pmb n}\cdot[\nabla'\psi_{in} +(ik-\frac{1}{r})\frac{{\pmb
r}}{r}\psi_{in}]ds,
\end{equation}
where $\psi_{out}({\pmb r},t)$ is the diffraction wave function on
display screen, $\psi_{in}({\pmb r},t)$ is the wave function of
slit surface ($z=c$) and $s$ is the area of the aperture or slit.

For the double-slit diffraction, Eq. (14) becomes
\begin{eqnarray}
\psi_{out}({\pmb r},t)&=&-\frac{1}{4\pi}\int_{s_{1}}
\frac{e^{ikr}}{r} {\pmb n}\cdot[\nabla'\psi_{1}
+(ik-\frac{1}{r})\frac{{\pmb r}}{r}\psi_{1}]ds \nonumber\\&&
-\frac{1}{4\pi}\int_{s_{2}}\frac{e^{ikr}}{r}{\pmb n}
\cdot[\nabla'\psi_{2} +(ik-\frac{1}{r})\frac{{\pmb
r}}{r}\psi_{2}]ds.
\end{eqnarray}
In Eq. (15), the first and second terms are corresponding to the
diffraction wave functions of the first slit and second slit.

   In the following, we firstly calculate the diffraction wave
function of the first slit, it is
\begin{equation}
\psi_{out_{1}}({\pmb
r},t)=-\frac{1}{4\pi}\int_{s_{1}}\frac{e^{ikr}}{r} {\pmb
n}\cdot[\nabla'\psi_{1} +(ik-\frac{1}{r})\frac{{\pmb
r}}{r}\psi_{1}]ds.
\end{equation}
The diffraction area is shown in FIG. 2, where
$k=\sqrt{\frac{2ME}{\hbar^{2}}}$, $s_{1}$ is the area of the first
single-slit, ${\pmb r'}$ is the position of a point on the surface
(z=c), $P$ is an arbitrary point in the diffraction area, and
${\pmb n}$ is a unit vector, which is normal to the surface of the
slit. \setlength{\unitlength}{0.1in}
 \begin{center}
\begin{figure}
\begin{picture}(100,10)
 \put(26,5){\vector(1,0){3}}
 \put(26,5){\vector(0,1){2.2}}
 \put(26,5){\vector(2,1){5}}

 \put(24,1){\line(1,0){2}}
 \put(24,1){\line(0,1){4}}
 \put(26,1){\line(0,1){4}}
 \put(24,5){\line(1,0){2}}

 \put(24,10){\line(1,0){2}}
 \put(24,10){\line(0,1){4}}
 \put(26,10){\line(0,1){4}}
 \put(24,14){\line(1,0){2}}

 \put(37.6,1){\line(0,1){15}}
 \put(26,5){\line(1,0){11.6}}

 \put(26,7){\line(3,1){11.5}}
 \put(26,7){\vector(3,1){5}}
 \put(26,5){\line(2,1){11.5}}
 \put(27.5,3.2){\makebox(2,1)[l]{${\pmb n}$}}
  \put(32,3.2){\makebox(2,1)[l]{${ l}$}}
 \put(24,7){\makebox(2,1)[c]{${\pmb r'}$}}
 \put(25,5){\makebox(2,1)[l]{$o$}}
 \put(31,6.3){\makebox(2,1)[c]{${\pmb R}$}}
 \put(30,9){\makebox(2,1)[c]{${\pmb r}$}}
 \put(38.5,10.5){\makebox(2,1)[l]{$P$}}
 \put(25,-0.5){\makebox(2,1)[l]{$c$}}
 \put(38.3,4){\makebox(2,1)[l]{${o'}$}}
 \put(38.3,6.5){\makebox(2,1)[l]{${S}$}}
\end{picture}
\caption{Diffraction area of the single slit} \label{moment}
\end{figure}
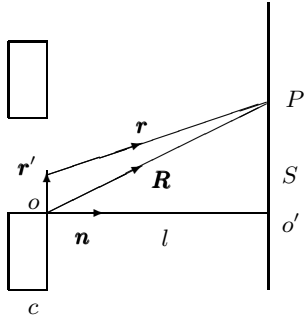
\end{center}
From FIG.\,2, we have
\begin{eqnarray}
r&=&R-\frac{{\pmb R}}{R}\cdot{\pmb r'}
\approx R-\frac{{\pmb r}}{r}\cdot{\pmb r'}\nonumber\\
 &=&R-\frac{{\pmb k_{2}}}{k}\cdot{\pmb r'},
\end{eqnarray}
and then,
\begin{eqnarray}
\frac{e^{ikr}}{r}&=&\frac{e^{ik(R-\frac{{\pmb r}}{r}\cdot{\pmb
r'})}} {R-\frac{{\pmb r}}{r}\cdot{\pmb r'}}
=\frac{e^{ikR}e^{-i{\pmb k_{2}}\cdot{\pmb r'}}}
{R-\frac{{\pmb r}}{r}\cdot{\pmb r'}}\nonumber\\
&\approx&\frac{{e^{ikR}e^{-i{\pmb k_{2}}\cdot{\pmb r'}}}}{R}
\hspace{0.3in}(|{\pmb r'}|\ll R),
\end{eqnarray}
with ${\pmb k_{2}}=k\frac{{\pmb r}}{r}$. Substituting Eq. (17) and
(18) into Eq. (16), one can obtain
\begin{eqnarray}
\psi_{out_{1}}({\pmb r},t)&=&-\frac{e^{ikR}}{4\pi
R}e^{-\frac{i}{\hbar}Et}\int_{s_{0}}e^{-i{\pmb k_{2}}\cdot {\pmb
r'}}\sum_{m=0}^{\infty}\sum_{n=0}^{\infty}\frac{16A}{(2m+1)(2n+1)\pi^{2}}
\nonumber\\&& \cdot
e^{i\sqrt{\frac{2ME}{\hbar^{2}}-(\frac{(2n+1)\pi}{b})^{2}-(\frac{(2m+1)\pi}{a_{1}})^{2}}\cdot
c} \sin \frac{(2n+1)\pi}{b}x'\sin
\frac{(2m+1)\pi}{a_{1}}y'\nonumber\\&& \cdot
[i\sqrt{\frac{2ME}{\hbar^{2}}-(\frac{(2n+1)\pi}{b})^{2}-(\frac{(2m+1)\pi}{a_{1}})^{2}}+
i{\pmb n}\cdot {\pmb k_{2}}-\frac{{\pmb n}\cdot {\pmb
R}}{R^{2}}]dx'dy'.
\end{eqnarray}
Assume that the angle between ${\pmb k_{2}}$ and $x$ axis ($y$
axis) is $\frac{\pi}{2}-\alpha$ ($\frac{\pi}{2}-\beta$), and
$\alpha (\beta)$ is the angle between ${\pmb k_{2}}$ and the
surface of $yz$ ($xz$), then we have
\begin{eqnarray}
k_{2x}=k\sin \alpha,\hspace{0.3in} k_{2y}=k\sin \beta,
\end{eqnarray}
\begin{eqnarray}
{\pmb n}\cdot {\pmb k_{2}}=k\cos \theta,
\end{eqnarray}
where $\theta$ is the angle between ${\pmb k_{2}}$ and $z$ axis,
and the angles $\theta$, $\alpha$, $\beta$ satisfy the equation
\begin{equation}
\cos^{2}\theta+\cos^{2}(\frac{\pi}{2}-\alpha)+\cos^{2}(\frac{\pi}{2}-\beta)=1.
\end{equation}

From FIG. 2, we have
\begin{equation}
sin\beta=\frac{s}{R},
\end{equation}
with $R=\sqrt{l^{2}+s^{2}}$. Substituting Eqs. (20)-(23) into Eq.
(19) yields
\begin{eqnarray}
\psi_{out_{1}}(x,y,z,t)&=&-\frac{e^{ikR}}{4\pi
R}e^{-\frac{i}{\hbar}Et}\sum_{m=0}^{\infty}\sum_{n=0}^{\infty}\frac{16A}{(2m+1)(2n+1)\pi^2}
e^{i\sqrt{\frac{2ME}{\hbar^{2}}-(\frac{(2n+1)\pi}{b})^{2}-(\frac{(2m+1)\pi}{a_{1}})^{2}}\cdot
c}\nonumber\\&& \cdot
[i\sqrt{\frac{2ME}{\hbar^{2}}-(\frac{(2n+1)\pi}{b})^{2}-(\frac{(2m+1)\pi}{a_{1}})^{2}}+(ik-\frac{1}{R})
\sqrt{\cos^{2}\alpha-(\frac{s}{R})^{2}}]\nonumber\\&&
\cdot\int^{b}_{0}e^{-ik\sin\alpha\cdot
x'}\sin\frac{(2n+1)\pi}{b}x'dx'\int^{a_{1}}_{0}e^{-ik\sin\beta\cdot
y'} \sin \frac{(2m+1)\pi}{a_{1}}y'dy'.
\end{eqnarray}
Equation (24) is the diffraction wave function of the first slit.
The neutron diffraction wave-function for the second slit can be
obtained by making the coordinate translations $x'=x, y'=y-(a+d),
z'=z$, it is
\begin{eqnarray}
\psi_{out_{2}}(x,y,z,t)&=&-\frac{e^{ikR}}{4\pi
R}e^{-\frac{i}{\hbar}Et}\sum_{m=0}^{\infty}\sum_{n=0}^{\infty}\frac{16A}{(2m+1)(2n+1)\pi^2}
e^{i\sqrt{\frac{2ME}{\hbar^{2}}-(\frac{(2n+1)\pi}{b})^{2}-(\frac{(2m+1)\pi}{a_{2}})^{2}}\cdot
c}\nonumber\\&&
[i\sqrt{\frac{2ME}{\hbar^{2}}-(\frac{(2n+1)\pi}{b})^{2}-(\frac{(2m+1)\pi}{a_{2}})^{2}}+(ik-\frac{1}{R})
\sqrt{\cos^{2}\alpha-(\frac{s}{R})^{2}}] \nonumber\\&&
\int^{b}_{0}e^{-ik\sin\alpha\cdot
x^{'}}\sin\frac{(2n+1)\pi}{b}x^{'}dx^{'} \nonumber\\&&
\int^{a_{1}+a_{2}+d}_{a_{1}+d}e^{-ik\sin\beta\cdot y^{'}} \sin
\frac{(2m+1)\pi}{a_{2}}(y^{'}-(a_{1}+d))dy^{'},
\end{eqnarray}
where $d$ is the two slit distance. The total diffraction wave
function for the double-slit is
\begin{eqnarray}
\psi_{out}(x,y,z,t)=c_{1}\psi_{out_{1}}(x,y,z,t)+c_{2}\psi_{out_{2}}(x,y,z,t),
\end{eqnarray}
where $c_{1}$ and $c_{2}$ are superposition coefficients , and
$|c_{1}|^{2}+|c_{2}|^{2}=1$. For the single-slit diffraction, we
can obtain the relative diffraction intensity $I$ on the display
screen,

\begin{equation}
I\propto|\psi_{out1}(x,y,z,t)|^{2}.
\end{equation}

For the double-slit diffraction, we can obtain the relative
diffraction intensity $I$ on the display screen,
\begin{eqnarray}
I&\propto&|\psi_{out}(x,y,z,t)|^{2}\nonumber\\&=&
{c_{1}^{2}}|\psi_{out1}(x,y,z,t)|^{2}+c_{2}^{2}|\psi_{out2}(x,y,z,t)|^{2}
+2c_{1}c_{2}Re[\psi_{out1}^*(x,y,z,t)\psi_{out2}(x,y,z,t)].
\end{eqnarray}.

\maketitle {\bf 4. Decoherence effect in double-slit
diffraction}\vskip 8pt

Decoherence is introduced here using a simple phenomenological
theoretical model that assumes an exponential damping of the
interferences [7, 18, 19], i.e., the decoherence is the dynamic
suppression of the interference terms owing to the interaction
between system and environment. The Eq. (26) describes the
coherence state coherence superposition, without considering the
interaction of system with external environment. When we consider
the effect of external environment, the total wave function of
system and environment for the double-slit factorizes as [7]
\begin{eqnarray}
\psi_{out}(x,y,z,t)=c_{1}\psi_{out_{1}}\otimes |E_{1}>_{t}
  +c_{2}\psi_{out_{2}}\otimes|E_{2}>_{t},
\end{eqnarray}
where $|E_{1}>_{t}$ and $|E_{2}>_{t}$ describe the state of the
environment. The diffraction intensity on the screen is now given
by[7]:
\begin{eqnarray}
I=(1+|\alpha_{t}|^{2})({c_{1}^{2}}|\psi_{out1}(x,y,z,t)|^{2}+c_{2}^{2}|\psi_{out2}(x,y,z,t)|^{2}
+2c_{1}c_{2}\Lambda_{t}Re[\psi_{out1}^*(x,y,z,t)\psi_{out2}(x,y,z,t)]).
\end{eqnarray}
where $\alpha_{t}=_{t}<E_{2}|E_{1}>_{t}$, and
$\Lambda_{t}=\frac{2|\alpha_{t}|}{1+|\alpha_{t}|^{2}}$. Thus,
$\Lambda_{t}$ is defined as the quantum coherence degree.  In Eq.
(30), the two slits wave functions $\psi_{out 1}$ and $\psi_{out
2}$ are calculated by the quantum approach (in Eqs. (24)-(25)). In
Refs. [7], the two slits wave functions are two Gaussian wave
packets. The fringe visibility of $\nu$ is defined as [7]:
\begin{equation}
\nu=\frac{I_{max}-I_{min}}{I_{max}+I_{min}},
\end{equation}
$I_{max}$ and $I_{min}$ being the intensities corresponding to the
central maximum and the first minimum next to it, respectively.
The value for the fringe visibility of $\nu=0.59$ is obtained in
Zeilinger et. al. experiment [20] ($I_{max}=4076, I_{min}=1050$),
and the quantum coherence degree $\Lambda_{t}\approx\nu$ [7].

\vskip 8pt {\bf 5. Numerical result} \vskip 8pt

Next, we present our numerical calculation of relative diffraction
intensity. The main input parameters are: neutron mass
$M=1.67\times10^{-27}$kg, the distance between slit and display
screen $l=5$m, the diffraction angle on $yz$ surface $\alpha=0$
rad, the slit thickness $c=3.0\times 10^{-5}$m, the neutron energy
$E=3.3\times 10^{-23}$J (corresponding to neutron wave length
$\lambda=20\AA$) and Planck's constant
$\hbar=1.055\times10^{-34}$Js. The equations (24)-(30) are series
for the integer $m$ and $n$. We find the series is convergence
when $m\geq600$ and $n\geq10$, so we can make numerical
calculation for equations (24)-(30). For single-slit experiment,
the neutron wavelength $\lambda=20\AA$, the slit width
$a_{1}=90\mu m$. In our calculation, we take the same experiment
parameters above, and the theoretical input amplitude parameter
$A=2.45\times 10^{4}$. From Eq. (27), we can obtain the
diffraction intensity pattern, and it is shown in FIG. 3. In FIG.
3, the solid curve is our calculation result, and the dot curve is
the experiment data [20]. From FIG. 3, we can find the calculation
result is agreement with experiment data. For the double-slit
diffraction, we consider two cases: coherence superposition and
decoherence effect. For the coherence superposition, we can
calculate the diffraction intensity from Eq. (28), and it is shown
in FIG. 4. The experiment parameters are: the neutron wavelength
$\lambda=20\AA$, the first and second slit width $a_{1}=21.9\mu
m$, $a_{2}=22.5\mu m$, the distance between the two slit $d=100\mu
m$. In our calculation, we take the same experiment parameters
above, and the theoretical input parameters are superposition
coefficients $c_{1}=0.397$, $c_{2}=0.918$
($|c_{1}|^{2}+|c_{2}|^{2}=1$) and amplitude parameter $A=6.8\times
10^{-2}$. In FIG. 4, the solid curve is our theoretical
calculation, and the dot curve is the experiment data [20]. From
the FIG. 4, we find that the theoretical result is in accordance
with the experiment data, when the position $s$ is in the range of
$|s|\geq 300 \mu m $. When the position $s$ is in the range of
$|s|\leq 300 \mu m $, the theoretical result has a large
discrepancy with the experiment data. We find the discrepancy can
be eliminated when the decoherence effect is considered. From Eq.
(30), we can obtain the diffraction intensity pattern and it is
shown in FIG. 5. In calculation, superposition coefficients
$c_{1}=0.397$, $c_{2}=0.918$, amplitude $A=6.8\times10^{-2}$ and
quantum coherence degree $\Lambda_{t}=0.59$. In FIG. 5, the solid
curve is our theoretical calculation, and the dot curve is the
experiment data [20]. From FIG. 5, we can find when the
decoherence effect is considered, the calculation result is in
accordance with the experiment data, and the discrepancy between
the theoretical result and experiment data can be eliminated.

\vskip 10pt

{\bf 6. Conclusion} \vskip 8pt In conclusion, we study neutron
single and double-slit diffraction with quantum theory approach.
The calculation result of single-slit diffraction is in accordance
with the experiment data. For the double-slit diffraction, we
study the diffraction intensity by two approaches, which are the
coherence superposition and decoherence mechanism. When we
consider the coherence superposition, the theoretical result has a
large discrepancy with the experiment data. When we consider the
decoherence mechanism, the theoretical result is in accordance
with the experiment data, and the discrepancy between the
theoretical result and experiment data has be eliminated.
Otherwise, we think the approach has universal applicability, such
as, it can also study electron, atom and molecular diffraction,
and it can also be studied multi-slit and grating diffraction.  \\

\newpage
\begin{figure}[tbp]
\begin{picture}(50,35)
{\resizebox{12cm}{8cm}{\includegraphics{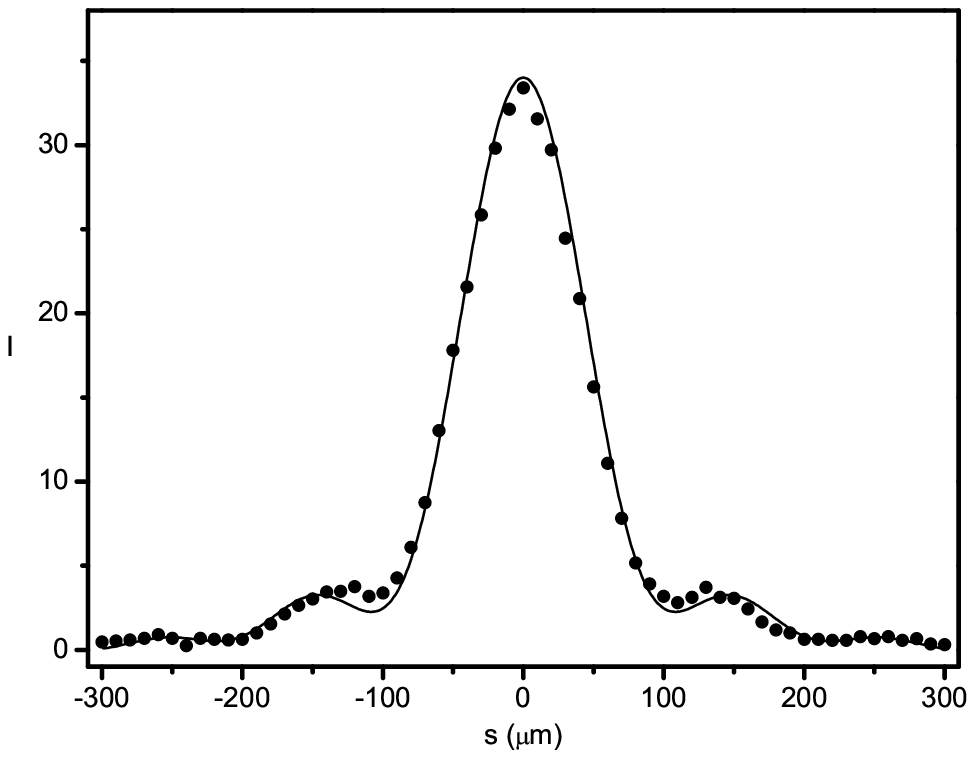}}}
\end{picture}
 \vskip 5pt
{\hspace{0.1in}FIG. 3: Comparison between theoretical prediction
from Eq. (27) (solid line) and experimental data taken from
[20](circle point) for neutron single-slit diffraction.}
 \label{moment}
\end{figure}

\begin{figure}[tbp]
\begin{picture}(50,40)
{\resizebox{12cm}{8cm}{\includegraphics{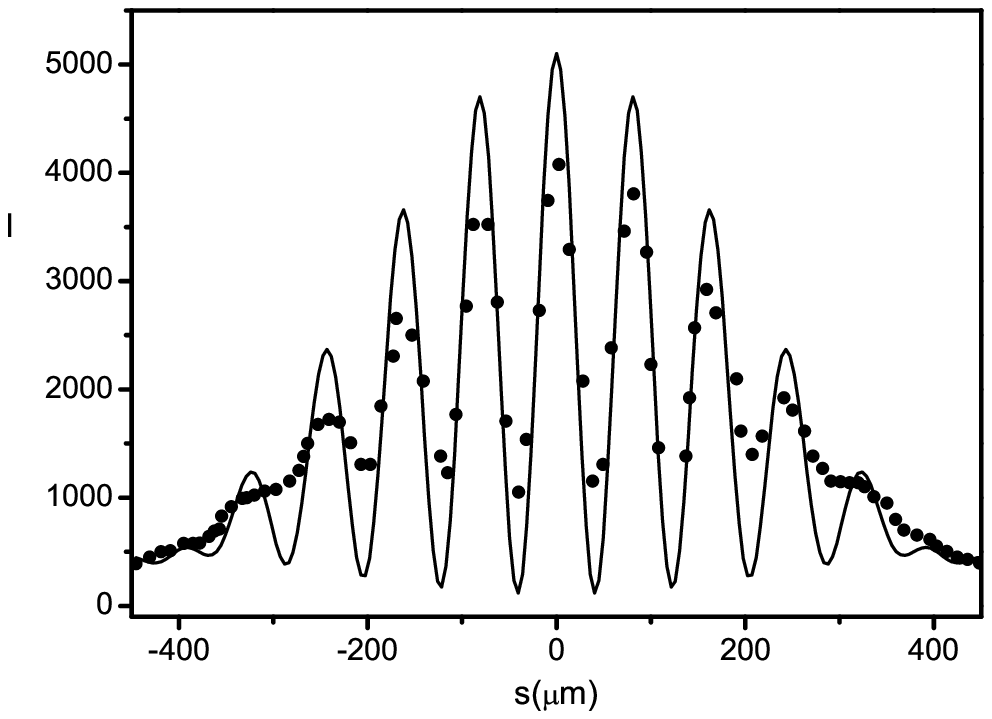}}}
\end{picture}
 \vskip 5pt
{\hspace{0.1in}FIG. 4: Comparison between theoretical prediction
from Eq. (28) (solid line) and experimental data taken from
[20](circle point) for neutron double-slit diffraction, no
including the decoherence effects.} \label{moment}
\end{figure}

\begin{figure}[tbp]
\begin{picture}(50,40)
{\resizebox{12cm}{8cm}{\includegraphics{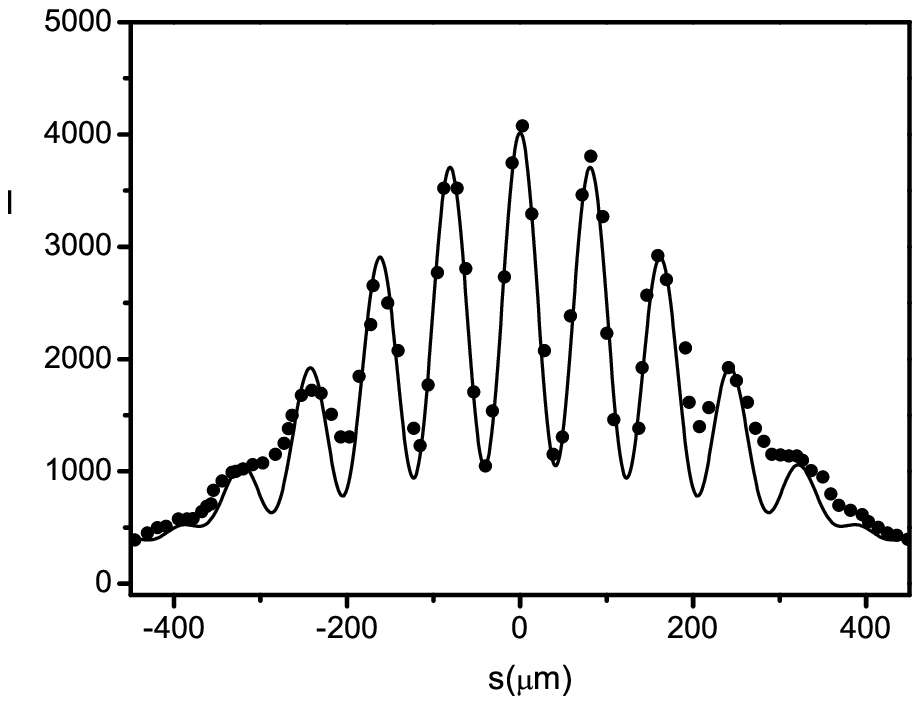}}}
\end{picture}
 \vskip 5pt
{\hspace{0.1in}FIG. 5: Comparison between theoretical prediction
from Eq. (28) (solid line) and experimental data taken from
[20](circle point) for neutron double-slit diffraction, no
including the decoherence effects.} \label{moment}
\end{figure}


\vskip 10pt
\newpage
\begin{thebibliography}{100}

\bibitem{s1}
O. Carnal and J. Mlynek, Phys. Rev. Lett. {\bf 66} 2689 (1991)
\bibitem{s2}
W. Sch\"{o}llkopf and P. J. Toennies, Science {\bf 266} 1345
(1994)
\bibitem{s3}
M. Arudt, O. Nairz, J. Voss-Andreae, C.  Kwller, G.  Vander Zouw
and A. Zeilinger, Nature {\bf 401} 680 (1999)
\bibitem{s4}
O. Nairz, M. Arudt and A. Zeilinger, J. Mod. Opt. {\bf 47} 2811
(2000)
\bibitem{s5}
Kunze S, Dieckmann k and Rempe G, Phys. Rev. Lett. {\bf 78} 2038
(1997)
\bibitem{s6}
B. Brezger, L. Hackermuller, S. Uttenthaler, J. Petschinka, M.
Arndt and A. Zeilinger, Phys. Rev. Lett. {\bf 88}, 100404 (2002)
\bibitem{s7}
S. A. Sanz, F. Borondo and J. M. Bastiaans, Phys. Rev. A {\bf 71},
042103 (2005)
\bibitem{s8}
O. Carnal and J. Mlynek, Phys. Rev. Lett. {\bf 66}, 2689 (1991)
\bibitem{s9}
W. Sch\"{o}llkopf and P. J. Toennies, Science {\bf 266}, 1345
(1994)
\bibitem{s10}
M. Arudt, O. Nairz, J. Voss-Andreae, C.  Kwller,  G.Vander Zouw
and A. Zeilinger, Nature {\bf 401}, 680 (1999)
\bibitem{s11}
O. Nairz, M. Arudt and A. Zeilinger, J. Mod. Opt. {\bf 47}, 2811
(2000)
\bibitem{s12}
S. Kunze, K. Dieckmann and G. Rempe, Phys. Rev. Lett. {\bf 78},
2038 (1997)
\bibitem{s13}
X. Y. Wu, B. J. Zhang, H. B. Li, J. B. Lu, X. J. Liu, L. Wang, C.
L. Zhang, B. Liu, X. H. Fan and Y. Q. Guo, Chin. Phys. Lett. {\bf
24}, 2741 (2007)
\bibitem{s14}
Xiang-Yao Wu, Bai-Jun Zhang and Xiao-Jing Liu Chun-Li Zhang, Bing
Liu and Yi-Qing Guo, FIZIKA B (Zagreb) 18, 4 (2009).
\bibitem{s15}
Li Wang, Bai-Jun Zhang, Zhong Hua, Ji Li, Xiao-Jing Liu and
Xiang-Yao Wu, Progress of Theoretical Physics 121, 685 (2009).
\bibitem{s16}
A. Viale, M. Vicari and N. Zanghi, Phys. Rev. A {\bf68}, 063610
(2003).
\bibitem{s17}
R. Tumulka, A. Viale and N. Zanghi, Phys. Rev. A {\bf75}, 055602
(2007)
\bibitem{s18}
C. Kiefer and E. Joos, Decoherence: Concepts and Examples, in
Quantum Future, eds. P. Blanchard and A. Jadczyk (Springer,
Berlin, 1998).

\bibitem{s19}
E. Joos and H.D. Zeh, Zeit. Phys. {\bf 59B}, 223 (1985).
\bibitem{s20}
Anton Zeilinger, Roland Gahler, C. G. Shull and Walter Mampe, Rev.
Mod. Phys. {\bf 60}, 1067 (1988)
\bibitem{s21}
J.D. Jackson 1999 {\it Classical Electrodynamics} (Chichester:
John Wiley-Sons) chap 10 p 579


\end{thebibliography}
\end{document}